\documentclass[conference]{IEEEtran}
\makeatletter
\def\ps@headings{%
\def\@oddhead{\mbox{}\scriptsize\rightmark \hfil \thepage}%
\def\@evenhead{\scriptsize\thepage \hfil \leftmark\mbox{}}%
\def\@oddfoot{}%
\def\@evenfoot{}}
\makeatother
\usepackage[hidelinks=true]{hyperref}
\usepackage{algorithm}
\usepackage{algpseudocode}
\usepackage{amsthm}

\usepackage{amsmath}
\allowdisplaybreaks[1]

\usepackage{graphicx}
\usepackage{stackrel}
\usepackage{aliascnt}
\usepackage{paralist}
\usepackage{subfig}

\usepackage{todonotes}

\newtheorem{theorem}{Theorem}
\newaliascnt{lemma}{theorem}
\newtheorem{lemma}[lemma]{Lemma}
\aliascntresetthe{lemma}
\newaliascnt{corollary}{theorem}
\newtheorem{corollary}[corollary]{Corollary}
\aliascntresetthe{corollary}

\newcommand{\abs}[1]{\left\vert#1\right\vert}
\newcommand{\set}[1]{\left\{#1\right\}}

\newcommand{\floor}[1]{\lfloor #1 \rfloor}

\newcommand{\alg}{ALG}

\newcommand{\goodput}{{\sc BDG}}
\newcommand{\dugoodput}{{\sc $d$-uBDG}}
\newcommand{\dbfifo}{{\sc $d$-bFIFO}}
\newcommand{\bfifo}{{\sc bFIFO}}
\newcommand{\pg}{{\sc PG}}
\DeclareMathOperator{\mpg}{PG}
\newcommand{\pglong}{{\sc ProactiveGreedy}}
\newcommand{\gdlong}{{\sc Greedy}}
\newcommand{\gdslong}{{\sc Greedy$_{\text{slack}}$}}
\newcommand{\gd}{{\sc G}}
\DeclareMathOperator{\mgd}{G}
\newcommand{\oplong}{{\sc Opportunistic}}

\begin{document}

\title{Bounded Delay Scheduling with Packet Dependencies}

\author{
\IEEEauthorblockN{
Michael Markovitch
and
Gabriel Scalosub
}
\IEEEauthorblockA{
   Department of Communication Systems Engineering\\
   Ben-Gurion University of the Negev\\
   Beer-Sheva 84105, Israel\\
   Email:
   {\tt markomic@post.bgu.ac.il, sgabriel@bgu.ac.il}
}
}

\maketitle


\begin{abstract}
A common situation occurring when dealing with multimedia traffic is having large data frames fragmented into smaller IP packets, and having these packets sent independently through the network. For real-time multimedia traffic, dropping even few packets of a frame may render the entire frame useless.
Such traffic is usually modeled as having {\em inter-packet dependencies}.
We study the problem of scheduling traffic with such dependencies, where each packet has a deadline by which it should arrive at its destination. Such deadlines are common for real-time multimedia applications, and are derived from stringent delay constraints posed by the application.
The figure of merit in such environments is maximizing the system's {\em goodput}, namely, the number of frames successfully delivered.

We study online algorithms for the problem of maximizing goodput of delay-bounded traffic with inter-packet dependencies, and use competitive analysis to evaluate their performance.
We present competitive algorithms for the problem, as well as matching lower bounds that are tight up to a constant factor.
We further present the results of a simulation study which further validates our algorithmic approach and shows that insights arising from our analysis are indeed manifested in practice.
\end{abstract}



\section{Introduction}
\label{sec:introduction}

A recent report studying the growth of real-time entertainment traffic in the Internet predicts that by 2018 approximately 66\% of Internet traffic in North America will consist of real-time entertainment traffic, and most predominantly, video streaming~\cite{sandvine13global}.
Such traffic, especially as video definition increases, is characterized by having large application-level data frames being
fragmented into smaller IP packets which are sent independently throughout the network.
For stored-video one can rely on mechanisms built into various layers of the protocol stack (e.g., TCP) that ensure reliable data transfer. However, for real-time multimedia applications such as live IPTV and video conferencing, these mechanisms are not applicable due to the strict delay restrictions posed by the application (such traffic is therefore usually transmitted over UDP). These restrictions essentially imply that retransmission of lost packets is in most cases pointless, since retransmitted packets would arrive too late to be successfully decoded and used at the receiving end.
Furthermore, the inability to decode an original dataframe once too many of its constituent packets have been dropped, essentially means that the resources used by the network to deliver those packets that did arrive successfully, have been wasted in vain.
Since network elements make their decisions on a packet-level basis, and are unaware of such dependencies occurring between packets corresponding to the same frame, such utilization inefficiencies can be quite common, as also demonstrated in experimental studies~\cite{boyce98packet}.

Some of the most common methods employed to deal with the hazardous effect of packet loss in such scenarios focus on trading bandwidth for packet loss; The sender encodes the data frames while adding significant redundancy to the outgoing packet stream, an approach commonly known as forward error correction (FEC). This allows the user to circumvent the effect of packet loss, at the cost of increasing the rate at which traffic is transmitted. This makes it possible (in some cases) to decode the data frame even if some of its constituent packets are dropped. However, increasing the bandwidth may be prohibitively costly in various scenarios, such as wireless access networks, network transcoders, and CDN headends. In such environments it is not recommended, nor even possible in many cases, to employ such solutions.

In this work we study mechanisms and algorithms that are to be implemented within the network, targeted at optimizing the usage of network resources (namely, buffer space and link bandwidth), when dealing with such delay-sensitive traffic.
Previous models presenting solutions for packet dependencies focused on managing a bounded-buffer FIFO queue, and mainly addressed the questions of handling buffer overflows (see more details in Section~\ref{sec:previous-work}).
We consider a significantly different model where each arriving packet has a {\em deadline} (which may or may not be induced by a deadline imposed on the data frame to which it corresponds). We assume no bound on the available buffer space, but are required to maximize the system's {\em goodput}, namely, the number of frames for which all of their packets are delivered by their deadline.\footnote{It should be noted that the objective of maximizing goodput (on the frame-level) is in most cases significantly different than the common concept of maximizing {\em throughput} (on the packet-level).}
This model better captures the nature of real-time video streaming, where a data frame must be successfully decoded in {\em real-time}, based on some permissible deadline by which packets should arrive, that still renders the stream usable.

We consider traffic as being {\em burst-bounded}, i.e., there is an upper bound on the number of packets arriving in a time-slot. This assumption does not restrict the applicability of our algorithms, since it is common for traffic (and especially traffic with stringent Quality-of-Service requirements) to be regulated by some token-bucket envelope~\cite{kurose11computer}.

We present several algorithms for the problem and use competitive analysis to show how close they are from an optimal solution. This approach makes our results globally applicable, and independent of the specific process generating the traffic. We further provide some lower bounds on the performance of any deterministic algorithm for the problem. Finally, we perform an extensive simulation study which further validates our results.


\subsection{System Model}
\label{sec:model}


We consider a time-slotted system where traffic consists of a sequence of unit-size {\em packets}, $p_1,p_2,\ldots$, such that packets are logically partitioned into {\em frames}. Each frame $f$ corresponds to $k$ of the packets, $p^f_1,\ldots,p^f_k \in \set{p_1,p_2,\ldots}$, where we refer to packet $p^f_{\ell}$ as the {\em $\ell$-packet} of frame $f$.
For every packet $p$ we denote its arrival time by $a(p)$, and we assume that the arrival of packets corresponding to frame $f$ satisfies $a(p^f_{\ell})\leq a(p^f_{\ell+1})$ for all $\ell=1,\ldots,k-1$.
We make no assumption on the relation between arrival times of packets corresponding to different frames.
Each packet $p$ is also characterized by a {\em deadline}, denoted $e(p)$, by which it should be scheduled for delivery, or else the packet {\em expires}. We assume $e(p) \geq a(p)$ for every packet $p$, and define the {\em slack} of packet $p$ to be $r(p)=e(p)-a(p)$.
For every time $t$ and packet $p$ for which $t \in [a(p),e(p)]$, if $p$ has not yet been delivered by $t$, we say $p$ is {\em pending at $t$}. we further define its {\em residual slack at $t$} to be $r_t(p)=e(p)-t$.%
\footnote{Note that this is a tad different from the model used in~\cite{kesselman04buffer} since we allow a packet to be scheduled also at time $e(p)=a(p)+r(p)$.}

We refer to an arrival sequence as being {\em $d$-uniform} if for every packet $p$ in the sequence we have $r(p)=d$.
We assume that $k \leq d$, which implies that any arriving frame can potentially be successfully delivered (e.g., if all other frames are ignored).
We further let $b$ denote the {\em maximum burst size}, i.e., for every time $t$, the number of packets arriving at $t$ is at most $b$.

The packets arrive at a queue residing at the tail of a link with unit capacity.
The queue is assumed to be empty before the first packet arrival.
In each time-slot $t$ we have three substeps:
\begin{inparaenum}[(i)]
\item {\em the arrival substep}, where the new packets whose arrival time is $t$ arrive and are stored in the queue,
\item {\em the scheduling/delivery substep}, where at most one packet from the queue is scheduled for delivery, and
\item {\em the cleanup substep}, where every packet $p$ currently in the queue which can not be scheduled by its deadline is discarded from the queue, either because $r_t(p)=0$, or because it belongs to a frame which has multiple pending packets at time $t$ and it is not feasible to schedule at least one of them by its deadline. Such packets are also said to expire at time $t$.
\end{inparaenum}

For every frame $f$ and every time $t$, if $f$ is not yet successful, but all of its packets that have arrived by $t$ are either pending or have been delivered, then $f$ is said to be {\em alive at $t$}. Otherwise it is said to have {\em expired}.
A frame is said to be {\em successful} if each of its packets is delivered (by its deadline).

The {\em Bounded-Delay Goodput problem} (\goodput) is defined as the problem of maximizing the number of successful frames. When traffic is $d$-uniform, we refer to the problem as the {\em $d$-uniform \goodput\ problem} (\dugoodput).

The main focus of our work is designing online algorithms for solving the \goodput\ problem.
An algorithm is said to be {\em online} if at any point in time $t$ the algorithm knows only of arrivals that have occurred up to $t$, and has no information about future arrivals.
We employ competitive analysis~\cite{SleatorT-85,Borodin-ElYaniv} to bound the performance of the algorithms.
We say an online algorithm \alg\ is $c$-competitive (for $c \geq 1$) if for every finite arrival sequence it produces a solution who's goodput is at least a $1/c$ fraction from the optimal goodput possible. $c$ is then said to be an upper bound on the {\em competitive ratio} of \alg.
As is customary in studies of competitive algorithms, we will sometimes assume the algorithms works against an {\em adversary}, which generates the input as well as an optimal solution for this input. This view is especially useful when showing lower bounds.
For completeness, we also address the {\em offline} problem where the entire arrival sequence is given in advance. In such offline settings the goal is to study the {\em approximation ratio} guaranteed by an algorithm, where an offline algorithm is an said to be an {$\alpha$-approximation algorithm} if for every finite arrival sequence the goodput of the solution it produces is always at least a fraction $1/\alpha$ of the optimal goodput possible.


\subsection{Our Contribution}
\label{sec:our-contribution}

In this paper we provide the initial study of scheduling delay-bounded traffic in the presence of packet dependencies. We initially provide some initial observations on the offline version of the problem, and then turn to conduct a thorough study of the problem with $d$-uniform traffic, i.e., where all packets have uniform delay $d$, burst sizes are bounded by $b$, and each frame consists of $k$ packets.

In the offline settings, we show that hardness results derived for the bounded-size FIFO queue model are applicable to our problem as well, which implies that it is NP-hard to approximate the problem to within a factor of $o(k/\ln k)$, and that a $(k+1)$-approximation exists.

In the Online settings we provide a lower bound of $\Omega(b^{k-1})$ on the competitive ratio of any deterministic online algorithm for the problem, as well as several other refined lower bounds for specific values of the system's parameters. We also design online deterministic algorithms with competitive ratio that asymptotically matches our lower bounds. This means that our algorithms are optimal up to a (small) constant factor.

We complement our analytical study with a simulation study which studied both our proposed algorithms, as well as additional heuristics for the problem, and also explores various algorithmic considerations in implementing our solutions. Our simulation results show that our proposed solutions are close to optimal, and also provide strong evidence that the performance exhibited by our algorithms in simulation closely follow the expected performance implied by our analysis.

Due to space constraints, some of the proofs are omitted, and can be found in~\cite{our-tech-report}.


\subsection{Previous Work}
\label{sec:previous-work}

The effect of packet-level decisions on the the successful delivery of large data-frames has been studied extensively in the past decades. Most of these works considered FIFO queues with bounded buffers and focused on discard decisions made upon overflows~\cite{ramanathan95enforcing}, as well as more specific aspects relating to video streams~\cite{awad02goodput,gurses05simple}.
This research thrust was accompanied by theoretical work trying to understand the performance of buffer management algorithms and scheduling paradigms, where the underlying architecture of the systems employed FIFO queues with bounded buffers. The main focus of these works was the design of competitive algorithms in an attempt to optimize some figure of merit, usually derived from Quality-of-Service objectives (see~\cite{goldwasser10survey} for a survey).
However, most of the works within this domain assumed the underlying packets are independent of each other, and disregarded any possible structure governing the generation of traffic, and the effect the algorithms' decisions may have on such frame-induced traffic.

Recently, a new model dealing with packet dependencies was suggested in~\cite{kesselman13competitive}. They assumed arriving packets are partitioned into frames, and considered the problem of maximizing the system's goodput. The main focus of this work was buffer management of a single FIFO queue equipped with a buffer of size $d$, and the algorithmic questions was how to handle buffer overflows, and they presented both competitive algorithms as well as lower bounds for this problem.
%
In what follows we refer to this problem as the {\em $d$-bounded FIFO problem} (\dbfifo).
Following this work, a series of works studied algorithms for various variants of the problem~\cite{emek12online,mansour11competitive,mansour12overflow,scalosub13buffer}.
Our model differs significantly from this body of work since in our model we assume no bounds on the available buffer size (as is more common in queueing theory models), nor do we assume the scheduler conforms with a FIFO discipline. More generally, we focus our attention on the task of deciding which packet to schedule, where each arriving packet has a deadline by which it should be delivered, as opposed to the question of how one should deal with overflows upon packet arrival when buffering resources are scarce.

Another vast body of related work focuses on issues of scheduling, and scheduling in packet networks in particular, in scenarios where packets have deadlines. Earliest-Deadline-First schedulilng was studied in various contexts, including OS process scheduling~\cite{silberschatz12operating}, and more generally in the OR community~\cite{pinedo12scheduling}.
Our framework is most closely related to~\cite{kesselman04buffer} which considers a packet stream where each packet has a deadline as well as a weight, and the goal is to maximizing the weight of packets delivered by their deadline.
They also consider relations between this model and the bounded-buffer FIFO queue model, and present competitive algorithms in both settings. These results are related to our discussion of the offline settings in Section~\ref{sec:offline}.
Additional works provided improved competitive online algorithms for this problem (e.g.~\cite{englert12considering,jez12online}). However, none of these works considered the settings of packet-dependencies, which is the main focus of our work.

\section{The Offline Settings}
\label{sec:offline}

In order to study the \dugoodput\ problem in the offline settings, it is instructive to consider the \dbfifo\ problem studied in~\cite{kesselman13competitive}. We recall that in this problem traffic arrives at a FIFO queue with buffer capacity $d$, and the goal is to maximize the number of frames for which all of their packets are successfully delivered (and not dropped due to buffer overflows).

In what follows we first prove that these two problems are equivalent in the offline settings (proof omitted).

\begin{lemma}
\label{lem:fifo-delay-equivalence}
For any arrival sequence $\sigma$, a set of frames $F$ constitutes a feasible solution to the \dugoodput\ problem if and only if it is a solution to the \dbfifo\ problem.
\end{lemma}

\begin{IEEEproof}
Assume a \dbfifo\ algorithm $A$, and a \dugoodput\ algorithm $B$, and note the set of packets in the queue of $A$ at time $t$ as - $P_{F}^{A}(t)$, and the set of packets in the buffer of $B$ at time $t$ as - $P_{U}^{B}(t)$.

At the time of arrival, every packet that a $A$ can choose to enqueue can be held in the buffer of $B$ (since there are no capacity constraints).
Every packet that $A$ enqueues can not stay in the queue more than $d$ time slots, since after $d$ time slots the packet must have been either sent or discarded (pre-empted) - the packet can not be in the queue longer than the slack time $d$. Therefore any algorithm $B$ which never schedules a packet before it is scheduled by $A$ can maintain that $P_{F}^{A}(t)\in P_{U}^{B}(t)$.

As at any time $t$ an $B$ can hold all the packets which $A$ can hold, any schedule which is feasible for $A$ algorithm, is also feasible for $B$ (including the optimal schedule).

For the reverse direction, assume that $B$ creates the schedule $S_{B}$. At any time $t$, of all the packets in the buffer at that time, no more than $d$ packets can be part of the schedule $S_{B}$ - if there were more than $d$ packets than not all of them could have been sent, rendering the schedule infeasible.

Therefore at any time $t$, all the buffered packets of the schedule $S_{B}$ can fit a FIFO queue of size $d$.
Since a packet can not stay in a FIFO queue and in the unbounded buffer more than $d$ time slots, there must exist an offline FIFO schedule $S_{A}$, for which all the packets in $S_{B}$ are in $S_{A}$.
\end{IEEEproof}


Note that in particular, \autoref{lem:fifo-delay-equivalence} implies that a set of frames $F$ is optimal for \dugoodput\ if and only if it is optimal for \dbfifo. By using the results of~\cite{kesselman13competitive} for the \dbfifo\ problem we obtain the following corollaries:

\begin{corollary}
\label{cor:offline:hardness}
It is NP-hard to approximate the \goodput\ problem to within a factor of $o(\frac{k}{\ln k})$ for $k \geq 3$, even for 0-uniform instances.
\end{corollary}
\begin{IEEEproof}
Since any $o(\frac{k}{\ln k})$ approximation would imply an approximation of the same factor for the 1-\bfifo\ problem, the result follows from~\cite[Corollary 2]{kesselman13competitive}.
\end{IEEEproof}

\begin{corollary}
\label{cor:offline:alg}
There is a deterministic $(k+1)$-approximation algorithm for the \dugoodput\ problem.
\end{corollary}
\begin{IEEEproof}
One can apply the algorithm G-OFF specified in~\cite{kesselman13competitive}. The result follows from~\cite[Theorem 3]{kesselman13competitive}.
\end{IEEEproof}

\section{The Online Settings}
\label{sec:online}

The offline settings studied in \autoref{sec:offline}, and the relation between the \dugoodput\ problem and the \dbfifo\ problem, give rise to the question of whether one should expect a similar relation to be manifested in the online settings. In this section we answer this question in the negative.

A first fundamental difference is due to the fact that in the \dugoodput\ problem the scheduler is not forced to follow a FIFO discipline. This means that the inherent delay of packets stored in the back of the queue which occurs in a FIFO buffer (unless packets are discarded upfront) can be circumvented by the scheduler in the \dugoodput\ problem, allowing it to take priorities into account.
Another significant difference between the two problems is that while in the \dbfifo\ problem discard decisions in case of buffer overflow must be made {\em immediately} upon overflow, in the \dugoodput\ problem such decisions can be somewhat delayed. Intuitively, the online algorithm in the \dugoodput\ problem has more time to study the arrivals in the near future, before making a scheduling decision, and thus enable it to make somewhat better decisions, albeit myopic. We note that this view is also used in~\cite{jez12online,englert12considering} in the concepts of provisional schedules and suppressed packets (we give more details of these features in \autoref{sec:algorithm-design:scheduling}).

\subsection{Lower Bounds}
\label{sec:online:lower-bounds}


In this section we provide several lower bounds for various ranges of our systems parameters. The main theorem is the following:

\begin{theorem} \label{thm:low-bound}
Any algorithm for the \dugoodput\ problem with burst size $b \geq 2d$ has competitive ratio $\Omega(b^{k-1})$.
\end{theorem}

\begin{figure}[htb]
\centering
\includegraphics[width=9cm]{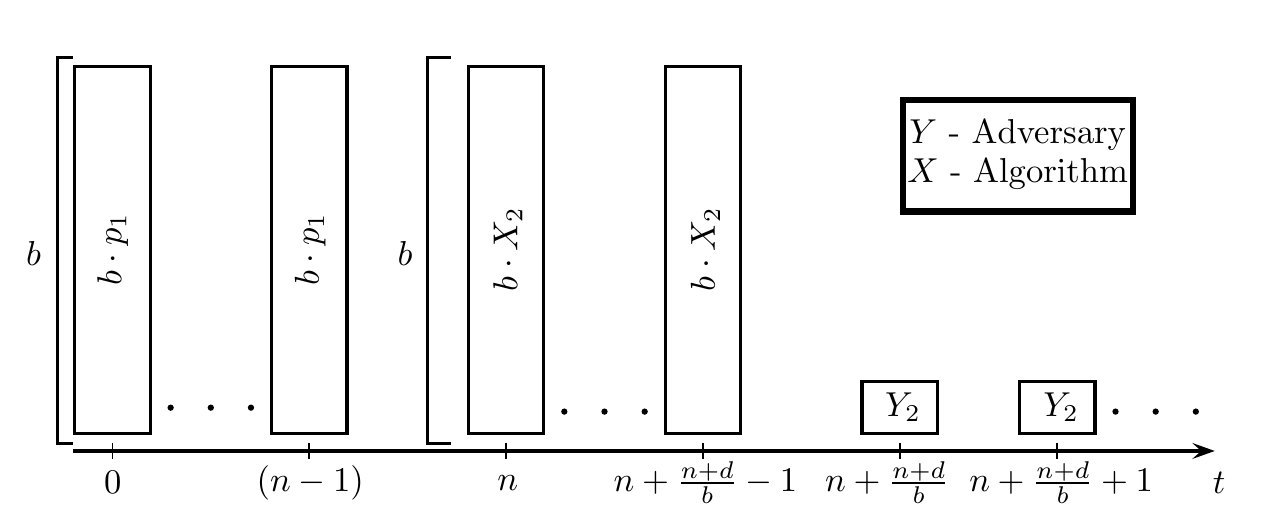}
\caption{Input for lower bound - The input is comprised of three stages. During the first stage the algorithm schedules $n+d$ '1' packets and the adversary schedules $n+d$ different '1' packets.
During the second stage the algorithm's '2' packets designated $X_{j}^{2}$ arrive as densely as possible. During stage 3 the adversary's '2' packets designated $Y_{j}^{2}$ arrive in a pattern that allows the adversary not to drop a single packet}
\label{lower-bound-input}
\end{figure}

\begin{IEEEproof}
Assume an arrival sequence with $b>1$, for traffic with slack $d$ comprised of three stages:

Stage 1 - at times $0, 1, 2, ..., (n-1)$, $b$ '1' packets arrive. During this stage, out of $nb$ '1' packets any online algorithm can only schedule up to $n+d$ '1' packets, and the adversary can schedule at least $n$ (and at most $n+d$) other packets - there are at most $d$ time slots for which the online algorithm can schedule all arriving packets. We choose $n$ so that $n+d$ is a multiple of $b$ in order to simplify the analysis.

Stage 2 - at times $n, n+1, ..., n+(n+d)/b-1$, $b$ '2' packets of frames whose '1' packets were scheduled by the online algorithm arrive. If we had not chosen $n+d$ to be a multiple of $b$, then there would also have been one more burst of size $n+d-\floor{(n+d)/b}$.

Stage 3 - Stage 3 - at times $n+(n+d)/b, n+(n+d)/b+1, ..., n+(n+d)/b+n(d-1)-d-1$, one '2' packet of frames whose '1' packet were not scheduled by the online algorithm arrive (including the adversary's packets).

This sequence is illustrated in \autoref{lower-bound-input}.

For $k>2$ stages 2 and 3 can be repeated with a slight modification to stage 2 - packets of frames which were scheduled by the online algorithm at the previous round arrive first (the burst contains frames whose packets were scheduled by the online algorithm in at least one stage).

By the end of stage 1, both the online and optimal algorithms would have sent $n+d$ '1' packets.
Stage 2 is designed to hit the goodput of the online algorithm as much as possible - only the online algorithm schedules packets.
Stage 3 is intended to maximize the goodput of the optimal algorithm.

The goodput of the optimal schedule for this input is at least $n$ (and at most $n+d$).
For k=2, the best possible goodput for an online unbounded buffer algorithm can not be better than $d+\frac{n+d}{b}$ - the algorithm can schedule up to $d$ packets of the last burst of stage 2, and for all earlier bursts of stage 2 no more than one packet each.
Since the at the end of the previous stage the number of frames whose '1' packets were scheduled by the algorithm is $n+d$, there are $\frac{n+d}{b}$ bursts in total.

If stages 2 and 3 are repeated, than the goodput of the optimal schedule for this input remains at least $n$.

In general, after stages 2 and 3 are performed $j$ times, the goodput ($GP$) of any online algorithm can not exceed $GP_j=d+\frac{GP_{j-1}}{b}$.

By induction: for $j=1$ we have from the definition of stage 1 that $GP_{j-1}=n+d$ and that $GP_j=d+\frac{GP_{j-1}}{b}$. For the induction stage consider $j+1$: during stage 2 there are $\frac{GP_j}{b}$ consecutive bursts (since packets scheduled during the previous round arrive first), and therefore the online algorithm can not schedule more than $\frac{GP_j}{b}+d$ packet, hence $GP_{j+1}=d+\frac{GP_j}{b}$.

The best possible goodput of an online algorithm with $k$ packets in a frame is then:
\begin{equation*}
\frac{n+d}{b^{k-1}}+\stackrel[i=2]{k}{\sum}\frac{d}{b^{i-2}}
\end{equation*}

Therefore, since we control $n$ and can make it as large as we desire ($nb$ is analogous to the number of streams), the lower bound for the competitive ratio of an online algorithm with a maximum burst size of $b$ is:

\begin{equation*}
\frac{|ALG|}{|OPT|}\leq\frac{n+d}{n\cdot b^{k-1}}+\stackrel[i=2]{k}{\sum}\frac{d}{n\cdot b^{i-2}}\xrightarrow[n\rightarrow\infty]{}\frac{1}{b^{k-1}}
\end{equation*}
\end{IEEEproof}

Our lower bound can be adapted to token-bucket regulated traffic, with maximum burst size $b$ and average rate $r$. Such restrictions on the traffic are quite common in SLAs. Of special interest is the case where the average rate is $r=1$, which essentially means the link is not oversubscribed. Even for such highly regulated traffic, we have the following lower bound:

%


\begin{theorem}
For token-bucket regulated traffic with parameters $(b,r=1)$, any algorithm for the \dugoodput\ problem where $b \geq 2d$ has competitive ratio $\Omega(\left(\frac{b}{d}\right)^{k-1})$.
\end{theorem}

\begin{IEEEproof}
This proof is very similar to the proof of \autoref{thm:low-bound}, and therefore we only give the differences.

The arrival sequence is modified so in stage 1 the interval between consecutive bursts is of length $2d$ (in order to build up for the next burst), and in stage 2 the interval between consecutive bursts is of length $b$. Stage 3 remains unchanged.

The goodput after repeating stages 2 and 3 $j$ times can not exceed $GP_j=d+\frac{d\cdot GP_{j-1}}{b}$, since out of every burst in stage 2 the online algorithm can schedule up to $d$ packets.

The best possible goodput for an online algorithm with $k$ packets in a frame is then:
\begin{equation*}
\frac{nd^k}{b^{k-1}}+\stackrel[i=2]{k}{\sum}\frac{d^{i-1}}{b^{i-2}}
\end{equation*}

Therefore, the lower bound for the competitive ratio of an online algorithm with a maximum burst size of $b$ is:

\begin{equation*}
\frac{|ALG|}{|OPT|}\leq\frac{nd^k}{nd\cdot b^{k-1}}+\stackrel[i=2]{k}{\sum}\frac{d^{i-1}}{nd\cdot b^{i-2}}\xrightarrow[n\rightarrow\infty]{}(\frac{d}{b})^{k-1}
\end{equation*}
\end{IEEEproof}
%


\subsection{The Proactive Greedy Algorithm}
\label{sec:proactive-greedy}

In this section we present a simple greedy algorithm, \pglong\ (\pg), that essentially ignores the deadlines in making scheduling decisions, and proactively drops packets from the queue.
Although one wouldn't expect such an algorithm to perform well in practice, its simplicity allows for a simple analysis which serves as the basis for the design and analysis of the refined greedy algorithm for the \dugoodput\ problem presented in subsequent sections.

For every time $t$ and frame $f$ that has pending packets at $t$, let $I_t(f)$ denote the index of the first pending packet of $f$. Recall that by our assumption on the order of packets within a frame, this is the minimal index of a pending packet corresponding to $f$.
We consider at every time $t$ all pending frames as ordered in decreasing order $(I_t(f)$.
For every packet $f$ we let $w(f)$ denote the number of packets corresponding to $f$ that were delivered by \pglong, i.e. $w(f)=\abs{\set{p \in f \mid p \text{ is delivered by \pg}}}$.
In what follows we slightly abuse notation and refer to a frame as alive as long as none of its packets has expired nor was dropped.
Algorithm \pglong\ is described in \autoref{alg:pg}.


\begin{algorithm}{}
\caption{\pglong: at the scheduling substep of time $t$}
\label{alg:pg}
\begin{algorithmic}[1]
\State drop all pending packets of frames that are not alive
\label{alg:pg:drop-dead-frames}
\State $Q_t \gets$ all alive frames with pending packets at $t$
\State $f \gets \arg\max_{f' \in Q_t} I_t(f')$
\Comment{Ties broken arbitrarily}
\label{alg:pg:identify-max-frame}
\State drop all pending packets of frames in $Q_t \setminus \set{f}$
\label{alg:pg:drop}
\State deliver the first pending packet of $f$
\label{alg:pg:schedule}
\end{algorithmic}
\end{algorithm}

The following lemma shows no packet ever expires in \pglong.

\begin{lemma}
\label{lem:no-expiry}
No packet ever expires in \pglong.
\end{lemma}

\begin{IEEEproof}
Consider some packet $p \in f$ for some frame $f$, and assume $p$ is not delivered. It follows that there exists some minimal time slot $t$ where $f$ ceases to be alive.
Consider time $a(p)$.
If $t<a(p)$, then $p$ is dropped upon arrival in line~\ref{alg:pg:drop-dead-frames}.
If $f$ is alive at $a(p)$ then $Q_{a(p)} \neq \emptyset$.
Let $f'$ be the frame identified in line~\ref{alg:pg:identify-max-frame}.
if $f \neq f'$, then $p$ is dropped in line~\ref{alg:pg:drop}, at time $a(p)$ and therefore does not expire.
Otherwise, we have $f=f'$.
Note that at the end of every scheduling substep the queue can only hold packets corresponding to the single frame identified in line~\ref{alg:pg:identify-max-frame} (if it is not empty).
It follows that $p$ is in the queue until time $t$, and since $d \geq k$, if it weren't dropped it could have been delivered successfully after all the preceding packets of $f$ residing in the queue with it at time $a(p)$, and wouldn't expire.
Since $p$ is not delivered, it follows that $p$ must be dropped at some time $t \leq a(p)+d$ due to some other frame $f'$ identified in line~\ref{alg:pg:drop}.
\end{IEEEproof}

The following corollary follows directly from \autoref{lem:no-expiry}.

\begin{corollary}
\label{cor:delivered-or-dropped}
Every frame in the arrival sequence is either successfully delivered by \pglong, or has one of its packets proactively dropped.
\end{corollary}


Let $F_{\mpg}$ be the set of frames successfully delivered by \pg, and let $O$ denote the set of frames successfully delivered by some optimal solution.

\begin{lemma}
\label{lem:mapping-target-exists}
If $f \notin F_{\mpg}$, then there exists a time $t=t_f$ such that $f$ is alive at $t$, packet $p_{I_t(f)} \in f$ is dropped in time $t$, and a packet $p' \in f'$ is delivered at time $t$, for some frame $f'$.
\end{lemma}
\begin{IEEEproof}
The proof follows directly from \autoref{lem:no-expiry}, and the details in its proof applied to packet $p_{I_t(f)}$ at the maximum time $t$ for which $f$ is alive at the beginning of time slot $t$.
\end{IEEEproof}

We describe a mapping $\phi$ of frames in the arrival sequence to frames in $F_{\mpg}$:
\begin{enumerate}
\item if $f \in F_{\mpg}$ then $f$ is mapped to itself.
\item if $f \notin F_{\mpg}$, then let $p_f$ be the first packet of $f$ dropped by \pg\ in line~\ref{alg:pg:drop}, and denote by $t_f$ the time slot where $p_f$ is dropped.
    Let $p'=p \in f'$ be the packet scheduled in time slot $t_f$ in line~\ref{alg:pg:schedule}.
    We map $f$ to $f'$ {\em directly}, and re-map any frames that were previously mapped to $f$ onto $f'$ {\em indirectly}.
    We say that these frames are re-mapped to $f'$ {\em via packet $p'$}.
    We also refer to $t_f$ as the {\em drop time of $f$} and to the set of frames mapped to $f$ via $p$ (either directly or indirectly) as $M(p)$.
\end{enumerate}

The following lemma shows that frames are remapped onto frames that are (strictly) closer to completion.

\begin{lemma}
\label{lem:mapping-weight-monotonicity}
If $f \notin F_{\mpg}$ is mapped to $f'$ then $w(f') > w(f)$.
\end{lemma}

\begin{IEEEproof}
Let $t$ be the drop time of $f$ and let $f'$ be the frame to which $f$ is directly mapped. By the choice of $f'$ in line~\ref{alg:pg:identify-max-frame} it follows that $I_t(f') \geq I_t(f)$. It follows that $w(f)=I_t(f)-1$, and since a packet of $f'$ is delivered in time $t$ we have $w(f') \geq I_t(f')$. Combining these inequalities we obtain
$$w(f') \geq I_t(f') \geq I_t(f) > I_t(f)-1 = w(f),$$
as required.
\end{IEEEproof}

The following corollary bounds the length of a re-mapping sequence.

\begin{corollary}
\label{cor:remapping-length}
A frame can be (re-)mapped at most $k$ times, and all frames are eventually mapped to frames in $F_{\mpg}$.
\end{corollary}

\begin{IEEEproof}
By definition every $f \in F_{\mpg}$ is mapped to itself.
For every frame $f \notin F_{\mpg}$, consider the number of times $\ell_f$ for which $f$ is mapped directly or indirectly to some other frame.
Denote by $f_{\ell}$ the $\ell$-th packet to which $f$ is mapped.
We prove by induction on $\ell$ that in the $\ell$-th such (re-)mapping, where $f$ is mapped to $f_\ell$, we have $w(f) < w(f)+\ell \leq w(f_{\ell})$.
This will imply that after at most $k$ remappings $f$ is mapped to a frame $f'$ for which $w(f')=k$, i.e., $f' \in F_{\mpg}$.
For the base case where $\ell=1$, this means $f$ is directly mapped to $f_1$. By \autoref{lem:mapping-weight-monotonicity} we have $w(f) < w(f_1)$, and therefore $w(f) + 1 \leq w(f_1)$.
For the induction step, consider the $\ell$-th remapping for $\ell>1$.
By the definition of the mapping, $f$ was mapped (directly or indirectly) to $f_{\ell-1}$ in the $(\ell-1)$-th remapping, and we are guaranteed to have $w(f)+(\ell-1) \leq w(f_{\ell-1})$.
By \autoref{lem:mapping-weight-monotonicity} we have $w(f_{\ell-1}) < w(f_{\ell})$, and therefore $w(f_{\ell-1}) + 1 \leq w(f_{\ell})$.
By combining the inequalities we obtain
$$w(f)+\ell = w(f) + (\ell-1) + 1 \leq w(f_{\ell-1}) + 1 \leq w(f_{\ell}),$$
thus completing the proof.
\end{IEEEproof}

The following corollary is an immediate consequence of \autoref{lem:mapping-target-exists} and \autoref{cor:remapping-length}.

\begin{corollary}
\label{cor:mapping-well-defined}
The mapping $\phi$ is well defined.
\end{corollary}

\begin{lemma}
\label{lem:direct-mapping-load}
For every frame $f$, the number of frames directly mapped to $f$ via packet $p \in f$ is at most $b$.
\end{lemma}
\begin{IEEEproof}
As there can be at most $b$ packets arriving at $t$, and all carrying over from $t-1$ correspond to a single frame, the number of frames mapped via $p(t)\in f$ is at most $b$.
\end{IEEEproof}

By \autoref{lem:direct-mapping-load} it follows that the overall number of frames directly mapped to any single frame $f$ is at most $kb$. Combining this with \autoref{cor:remapping-length} implies a $(k\cdot b)$-ary depth-$k$ tree structure for the mapping (direct or indirect) onto any single frame $f \in F_{\mpg}$, which shows that \pglong\ delivers at least a fraction of $\frac{1}{(k\cdot b)^k}$ of the total arriving traffic.
This clearly serves as a bound on the competitive ratio.
However, a significantly better bound can be obtained by a closer examination of direct mappings.


\begin{lemma}
\label{lem:mapping-load}
For every frame $f$, the overall number of frames mapped to $f$ at time $t$ via packets $p_{\ell} \in f$ is at most \hbox{$b\cdot (1+b)^{\ell -1}$}.
\end{lemma}
\begin{IEEEproof}
First we observe that if a a frame $f'$ is directly mapped to a frame $f$ at time $t$, then $I_t(f') \leq I_t(f)$. In particular, if the minimal-indexed packet of $f'$ dropped at time $t$ is the $j$-th packet of $f'$, then $j \leq I_t(f)$.  Also notice that $(1+b)^{\ell-1}=\sum_{i=0}^{\ell-1}\binom{\ell-1}{i}b^i$.

We now turn to prove the claim by induction on $\ell$.
For the base case of $\ell=1$, assume $f'$ is mapped to $f$ via $p_{\ell}$ at time $t$. If $f'$ is mapped to $f$ directly, by the above observation we have that the minimal-indexed packet of $f'$ dropped at $t$ is at most $\ell=1$, and therefore it must be the first packet of $f'$. Since all these packets must have arrived at time $t$, it follows that none of these frames have any frames mapped to them. By \autoref{lem:direct-mapping-load} it follows that the overall number of frames directly mapped to $f$ via $p$ is at most $b$.
Note that this implies that for the base case there can be no frames indirectly mapped to $f$. It follows that the overall number of frames mapped to $f$ via $p_{1}$ is at most $|M(p_1)|=b=b\cdot (1+b)^0$, thus completing the base case.
For the induction step consider $p_{\ell+1} \in f$ for $\ell+1$, and let $f''$ be a frame mapped to $f$ via $p_{\ell+1}$.
Assume $f''$ is mapped to $f$ directly.
By the above observation we have that the minimal-indexed packet of $f''$ dropped at $t$ is at most $\ell+1$. Again, the overall number of frames directly mapped to $f$ via $p_{\ell+1}$ is at most $b$. It follows that the maximum index of a packet $p'' \in f''$ for which there were frames mapped to $f''$ via $p''$ is at most $\ell$.
Hence for $\ell+1$, the overall number of frames mapped to $f$ at time $t$ via $p_{\ell+1}^{f}$ is at most:

\begin{align}
|M(&p_{\ell +1})| = b\left(1+\sum_{i=1}^{\ell}|M(p''_{i})|\right)  \label{eq:binom:definition}\\
&\leq b\left(\binom{\ell}{0}\cdot b^0+\sum_{i=1}^{\ell}\sum_{j=0}^{i-1}b\cdot \binom{i-1}{j} \cdot b^{j}\right) \label{eq:binom:induction}\\
&= b\left(\binom{\ell}{0}\cdot b^0+b\sum_{i=1}^{\ell}b^{i-1}\sum_{j=0}^{\ell-i}\binom{i-1+j}{i-1}\right) \label{eq:binom:substitution} \\
&=b\left(\binom{\ell}{0}\cdot b^0+b\sum_{i=1}^{\ell}\binom{\ell}{i}b^{i-1}\right) \label{eq:binom:diagonal} \\
&= b\sum_{i=0}^{\ell}\binom{\ell}{i}b^{i}. \notag
\end{align}

Equality~\eqref{eq:binom:definition} follows from the direct mappings via $p_{\ell+1}$ and inequality~\eqref{eq:binom:induction} follows from the induction hypothesis. Equality~\eqref{eq:binom:substitution} follows from reversing the order of summation on $j$, and noticing that only the topmost $\ell-(i-1)$ sums over $j$ contribute to the coefficient of $b^{i-1}$. Finally, equality~\eqref{eq:binom:diagonal} is a simple diagonal binomial identity.

Since $f''$ itself is mapped to $f$ in addition to all the frames which were mapped to $f''$.
\end{IEEEproof}

Recall $O$ denotes the set of frames in an optimal solution. The following corollary provides a bound on the number of frames in $O \setminus F_{\mpg}$ that are mapped by our mapping procedure.

\begin{corollary}
\label{cor:mapping-load-of-opt}
For every frame $f$, the overall number of frames in $O \setminus F_{\mpg}$ mapped to $f$ at time $t$ via packets $p_{\ell} \in f$ is at most \hbox{$\min\set{d,b} \cdot\left(1+b\right)^{\ell-1}$}.
\end{corollary}
\begin{IEEEproof}
Assume a frame $f$ which has a packet $p\in f$ delivered by \pglong\ at time $t$.
The number of frames in $O \setminus F_{\mpg}$ which can be directly mapped to $f$ via $p\in f$ is at most $\min\set{d,b}$, since the optimal solution cannot deliver more than $d$ of the pending packets at any time $t$.

If $d\geq b$, then since the maximal number of packets which can arrive at time $t$ is $b$, the result of \autoref{lem:mapping-load} applies in this case too, and $b=\min\set{d,b}$.

If $d<b$, if frames $f'\in O \setminus F_{\mpg}$ were to be mapped to another frame $f''\in O \setminus F_{\mpg}$ via $p_{\ell}^{''}$ then at most $d-1$ frames with weight $w(f'_{\ell-1})$ or $d$ frames with weight $w(f'_{\ell-2})$ can be directly mapped via $p_{\ell}^{''}$.
Therefore, the highest number of frames $f'\in O \setminus F_{\mpg}$ mapped to a single frame $f\in F_{\mpg}$ is achieved when frames $f'\in O \setminus F_{\mpg}$ can not be mapped to frames $f''\in O \setminus F_{\mpg}$ (otherwise the resulting tree like structure contains less mapped frames) - the maximal number of frames $f'\in O \setminus F_{\mpg}$ mapped to a single frame $f\in F_{\mpg}$ is achieved when \pglong\ never schedules a packet of a frame $f'\in O \setminus F_{\mpg}$.
Hence by counting the maximal number of frames mapped to $f\in F_{\mpg}$ via packets with weight $w\geq 1$ according to \autoref{lem:mapping-load} ($b\left(1+b\right)^{\ell-2}$ mapped frames), and mapping at most $d$ frames $f'\in O \setminus F_{\mpg}$ via every scheduled packet with $\ell =1$ (for every frame mapped to $f$ including itself), the result for the maximal number of frames $f'\in O \setminus F_{\mpg}$ mapped to a single frame $f\in F_{\mpg}$ is $\left(b\left(1+b\right)^{\ell-2}+1\right)d\leq d\left(1+b\right)^{\ell-1}$. 
\end{IEEEproof}


\begin{theorem}
\label{thm:pg-competitive-ratio}
Algorithm \pglong\ is $O(\min\set{d,b} b^{k-1})$-competitive.
\end{theorem}
\begin{IEEEproof}
By \autoref{cor:mapping-load-of-opt} the overall number of frames in $O \setminus F_{\mpg}$ mapped to any $f \in F_{\mpg}$ is
\begin{align}
\sum_{\ell=1}^{k} \abs{M(p^f_{\ell})}
&\leq \min\set{d,b} \sum_{\ell=1}^{k} (1+b)^{\ell-1} \notag \\
&= \min\set{d,b} b^{k-1} (1 + O(\frac{k}{b})) \notag \\
&= O(\min\set{d,b} b^{k-1}), \notag
\end{align}
which completes the proof.
\end{IEEEproof}


From this analysis of the \pglong\ algorithm we learn that choosing a preference based on how close is a frame to completion guarantees not only that a frame will be completed (\autoref{cor:remapping-length}), but also that the algorithm will be competitive.
Also, even though this algorithm is very simple (conceptually), the competitiveness is close to the lower bound we proved for the general case (within a factor of $d$ from the lower bound). This competitive ratio does not depend if the traffic is burst bound ($r=b$ the general case) or token bucket shaped ($r<b$), since an arrival sequence achieving this bound can be created regardless of the value of $r$ (due to step 2 of the algorithm).

\subsection{The Greedy Algorithm}
\label{sec:greedy}

With the proactive greedy algorithm, we saw that choosing a preference based on how close is a frame to completion results in an upper bound which is close to the lower bound for the general case ($r=b$).
However, the \pglong\ algorithm is not a natural algorithm to suggest since frames are being dropped unnecessarily, resulting both in inefficiency and in implementation complexity.

We suggest a more intuitive algorithm, the \gdlong\ algorithm - \autoref{alg:greedy-algorithm}.
The only difference between the two algorithms is that the \gdlong\ algorithm does not drop frames unnecessarily - a frame expires only if it is not feasible to schedule one of it's packets by the packet's deadline.

\begin{algorithm}{}
\caption{\gdlong: at the scheduling substep of time $t$}
\label{alg:greedy-algorithm}
\begin{algorithmic}[1]
\State drop all pending packets of frames that are not alive
\label{alg:gd:drop-dead-frames}
\State $Q_t \gets$ all alive frames with pending packets at $t$
\State $f \gets \arg\max\set{ I_t(f') \mid f' \in Q_t}$
\label{alg:gd:identify-max-frame}
\State deliver the first pending packet of $f$
\label{alg:gd:schedule}
\end{algorithmic}
\end{algorithm}

We say a packet $p_{\ell}^f$ is eligible if at a time $t$ it is in the buffer of \gdlong\ and it's index is $\ell =I_{t}(f)$ - it is the first pending packet of the live frame $f$ at time $t$.
The following lemma relates the number of packets of index at least $\ell$ that were dropped by \gdlong, to the number of packets of index at least $\ell$ that were delivered by $t$.

\begin{lemma}
\label{lem:gd-direct-mapping-load}
For any time $t$ during which a previously eligible packet $p_\ell \in f$ is dropped, assume $n_t^{\ell}$ is the total number of packets with index at least $\ell$ which were eligible and dropped by time $t$. It follows that at least $\lceil n_t^{\ell}/b\rceil$ packets with index of at least $\ell$ were delivered by \gdlong by time $t$.
\end{lemma}

\begin{IEEEproof}
At most $b+1$ packets can become eligible at the start of any time $t$ - a burst of at most $b$ packets, and one packet $p'''_{i}\in f'''$ which was in buffer at time $t-1$ if packet $p'''_{i-1}\in f'''$ was scheduled at time t-1.

When a packet $p_{\ell}\in f$ is eligible, only a packet which belongs to a frame of at least the same weight can be scheduled. We note two possible cases:
\begin{enumerate}
\item For the case that $b+1$ packets belonging to frames with weight $w(f')\geq w(f)$ become eligible at time $t$ (including $p_{\ell}$), then one of them will be scheduled since all eligible packets carrying over from time $t-1$ $p''\in f''$ belong to frames with $w(f'')< w(f''')$ - at least one packet out of $b+1$ will be scheduled at time $t$.
\item For the case that up to $b$ packets belonging to frames with weight $w(f')\geq w(f)$ become eligible at time $t$ (including $p_{\ell}$), then one packet $p_{i\geq \ell}$ will be scheduled. The packet scheduled at time $t$ can be one of the up to $b$ packets that became eligible, or it can be a packet that arrived at an earlier time (and is still eligible) - at least one packet out of $b+1$ (in case the scheduled packet was already eligible at $t-1$) packets will be scheduled at time $t$.
\end{enumerate}

Note that if at time $t-1$ there were eligible packets with $w(f')\geq w(f)$, all of them were already accounted for at time of first eligibility - either during a previous case 2, or during a previous case 1.

The combination of the two cases guarantees that at any time $t$, if the number of packets with index of at least $\ell$ which were eligible and subsequently dropped by \gdlong\ by time $t$ is $n$, than at least $\lceil\frac{n}{b} \rceil$ packets with an index of at least $\ell$ were scheduled by time $t$.
\end{IEEEproof}

In order to find the upper bound of the competitive ratio, we use the same approach used to analyze \pglong.
Let $F_{\mgd}$ be the set of frames successfully delivered by \gdlong\, and let $O$ denote the set of frames successfully delivered by some optimal solution. We define mapping $\psi$ of frames in the arrival sequence to packets in $F_{\mgd}$.
\begin{enumerate}
\item if $f \in F_{\mgd}$ then $f$ is mapped to itself.
\item if $f\in O\setminus F_{\mgd}$, then let $p_{\ell}^f$ be the first packet of $f$ dropped by \gd\ in line~\ref{alg:gd:drop-dead-frames}, and denote by $t_f$ the time slot where $p_{\ell}^f$ was dropped.
    Let $p'_{i\geq\ell}=p(t\leq t_f) \in f'$ be a packet scheduled in time slot $t\leq t_f$ in line~\ref{alg:pg:schedule}.
    We map $f$ to $f'$ {\em directly}, and re-map any frames that were previously mapped to $f$ onto $f'$ {\em indirectly}, if it is the earliest scheduled packet with the lowest index $i\geq\ell$ which has less than $2$ packets already {\em directly} mapped to it.
    We say that these frames are re-mapped to $f'$ {\em via packet $p'=p(t\leq t_f)$}.
    We also refer to the group of all frames mapped to $f$ via $p$ as $M(p)$.
\item if $f \notin F_{\mgd}\cup O$, then let $p_{\ell}^f$ be the first packet of $f$ dropped by \gd\ in line~\ref{alg:gd:drop-dead-frames}, and denote by $t_f$ the time slot where $p_{\ell}^f$ was dropped.
    Let $p'_{i\geq\ell}=p(t\leq t_f) \in f'$ be a packet scheduled in time slot $t\leq t_f$ in line~\ref{alg:pg:schedule}.
    We map $f$ to $f'$ {\em directly}, and re-map any frames that were previously mapped to $f$ onto $f'$ {\em indirectly}, if it is the earliest scheduled packet with the lowest index $i\geq\ell$ which has less than $b$ packets already {\em directly} mapped to it.
    We say that these frames are re-mapped to $f'$ {\em via packet $p'=p(t\leq t_f)$}.
    We also refer to the group of all frames mapped to $f$ via $p$ as $M(p)$.
\end{enumerate}

The following lemma shows that the mapping is well defined, and that at most 2 packets of $O$ are directly mapped to any single frame $f$.

\begin{lemma}
\label{lem:direct-mapping-adversary-greedy}
At any time $t$ when there is no eligible packet with index $j\geq\ell$, at least 1 packet with index $i\geq\ell$ have been scheduled by \gdlong\ for every 2 packets with indices $j\geq\ell$ which were eligible and subsequently dropped by \gdlong.
\end{lemma}

\begin{IEEEproof}
Note that during every time slot that packet $p_{\ell}\in f \in O$ is eligible for \gdlong, if it is not scheduled it means there exist some packet $p'_{i\geq\ell}\in f'$ which is scheduled instead. Also note that the schedule of the adversary must be feasible.

Consider the first time a packet with index $\ell$ becomes eligible $t_e$: either all the packets with index $i\geq \ell$ which became eligible at $t_e$ (at least one of them with index $i=\ell$) arrived at $t_e$, or one of the packets with an index $i=\ell$ which became eligible was already in the buffer and became eligible because the $\ell -1$ packet of the same stream was scheduled at time slot $t_e-1$. 

We define $t_b$ as the first time after $t_e$ where there are no eligible packets with an index $i\geq \ell$ in the buffer, by definition $t_b>t_e$. We say call the interval $[t_e,t_b]$ an $\ell$-ary busy period.

During an $\ell$-ary busy period, at most $t_b-t_e+1$ of the adversary's packets with index $i\geq \ell$ can become eligible for \gdlong:
\begin{itemize}
\item at time $t_e$ up to $d+1$ of the adversary's packets with index $i\geq \ell$ can become eligible for \gdlong due to feasibility of $O$, as at most $d$ such eligible packets can arrive at time $t_e$ and one packet was already in the buffer can become eligible.
\item during the interval $[t_e,t_b]$ at most $t_b-t_e$ of the adversary's packets with index $i\geq \ell$ could arrive and become eligible for \gdlong, due to the combination of the feasibility of $O$ and of the fact that at time $t_b$ there are no eligible packets with index $i\geq \ell$ in the buffer.
\end{itemize}

During the $\ell$-ary busy period $[t_e,t_b]$, $t_b-t_e$ packets with index $i\geq \ell$ are scheduled by \gdlong.

Therefore during the $\ell$-ary busy period $[t_e,t_b]$, \gdlong\ schedules $t_b-t_e$ packets with index $i\geq \ell$, and drops at most $t_b-t_e+1$ of the adversary's packets with index $i\geq \ell$ which became eligible during the interval.

We extend the definition of $t_e$ to be the first time slot a packet with index $\ell$ becomes eligible after some time $t<t_e$ during which there were no eligible packets with index $i\geq \ell$ in the buffer (this definition applies also for $t_e=0$ as at previous times the buffer was empty). Then an $\ell$-ary busy period can be followed by another $\ell$-ary busy period after a it ends (there can be no overlap between the busy periods). Busy periods of different indices can and do overlap one another.

As all packets of index $\ell$ become eligible during an $\ell$-ary busy period the result follows, since during a single busy period $t_b-t_e$ packets with index $i\geq \ell$ are scheduled by \gdlong, and at most $t_b-t_e+1$ of eligible adversary's packets with index $i\geq \ell$ are dropped by \gdlong.

\end{IEEEproof}

\autoref{lem:gd-direct-mapping-load} and \autoref{lem:direct-mapping-adversary-greedy} ensure that mapping $\psi$ is well defined. The definition of $\psi$ together with \autoref{cor:remapping-length} yields:

\begin{corollary}
\label{cor:remapping-length-greedy}
A frame can be (re-)mapped at most $k$ times, and all frames are eventually mapped to frames in $F_{\mpg}$.
\end{corollary}

By the definition of $\psi$, no more than $b$ frames can be directly mapped to a frame $f'$ via packet $p'\in f'$. Furthermore, if $f\notin F_{\mgd}$ is mapped to $f'$ then $w(f')>w(f)$. It follows that the proof of \autoref{lem:mapping-load} also holds for \gdlong (since all requirements are met):

\begin{corollary}
\label{cor:gd-mapping-load}
For every frame $f$, the overall number of frames mapped to $f$ at time $t$ via packets $p_{\ell} \in f$ is at most \hbox{$b\cdot (1+b)^{\ell -1}$}.
\end{corollary}

Therefore by applying \autoref{cor:gd-mapping-load} and the fact that according to $\psi$ the number of frames in $O \setminus F_{\mpg}$ directly mapped to any $f$ via $p\in f$ is at most $2$, using similar arguments as the ones used in \autoref{cor:mapping-load-of-opt} and \autoref{thm:pg-competitive-ratio} we obtain the following theorem.

\begin{theorem}
\label{thm:gd-competitive-ratio}
Algorithm \gdlong\ is $O(b^{k-1})$-competitive.
\end{theorem}

\begin{IEEEproof}
Applying the arguments from \autoref{cor:mapping-load-of-opt} on the results of \autoref{cor:gd-mapping-load} \autoref{lem:direct-mapping-adversary-greedy}, yields that the overall number of frames in $O \setminus F_{\mpg}$ mapped to $f$ at time $t$ via packets $p_{\ell} \in f$ is at most \hbox{$2\cdot\left(1+b\right)^{\ell-1}$} (since $\min\set{2,b}\leq 2$).
Then as in \autoref{thm:pg-competitive-ratio} the overall number of frames in $O \setminus F_{\mpg}$ mapped to any $f \in F_{\mpg}$ is
\begin{align}
\sum_{\ell=1}^{k} \abs{M(p^f_{\ell})}
&\leq 2 \sum_{\ell=1}^{k} (1+b)^{\ell-1} \notag \\
&= 2 b^{k-1} (1 + O(\frac{k}{b})) \notag \\
&= O( b^{k-1}) \notag
\end{align}
\end{IEEEproof}

We note that by our lower bounds algorithm \gdlong\ optimal up to a constant factor. Furthermore it should be noted that this improved bound, as well as expected performance in practice, comes at a cost of significantly more complex implementation. Specifically, while \pglong\ can potentially be implemented using a FIFO buffer, \gdlong\ does not deliver packets in FIFO order. 
In other aspects \gdlong\ is not significantly more complex than the \pglong\, e.g., by trading garbage collection with killing live frames.


%
%

\section{Further Algorithmic Considerations}
\label{sec:algorithm-design}

\subsection{Tie-breaking}
\label{sec:algorithm-design:tie-breaking}

The results from the analysis of the two algorithms, show us that an algorithm should prefer frames which are closer to completion (since this characteristic guarantees competitiveness), and that live frames should be kept in the buffer as long as possible (as shown by the difference between the algorithms).
But the analysis brings up the question how to best tie-break between frames which are the same distance from completion (the number of sent packets is the same for both frames).

A natural choice for a tie-breaker is the residual slack $r_t(p^f)$ of the smallest-index packet $p^f \in Q_t \cap f$ for each frame $f$ that has the maximal $I_t(f)$ value.
The purpose of such a tie breaker is of course to improve performance by keeping as many frames alive as possible.

A less obvious choice for a tie-breaker is the number of pending packets corresponding to frame $f$, denoted $n_t(f)$.
The intuition underlying this choice is that preferring frames with lower $n_t(f)$ can more rapidly ``clear'' the effect of $f$ on other frames with pending packets.

One should note that neither choice affects the asymptotic competitiveness of \gdlong, which is tight up to a constant factor.
However, this choice is expected to influence the performance of the algorithms in practice.
In \autoref{sec:simulations} we further address these design dilemmas.

\subsection{Scheduling}
\label{sec:algorithm-design:scheduling}


For both of the greedy algorithms presented in Sections~\ref{sec:proactive-greedy} and~\ref{sec:greedy}, the first packet of the preferred frame was sent, where the difference between the algorithms boiled down to the the way other pending packets were treated.
In particular, the residual slack of the packets is essentially ignored by these greedy approaches (although it can be taken into account in tie-breaking, as discussed above).


One common approach to incorporate residual slack into the scheduler is considering {\em provisional schedules}, which essentially try pick the packet to be delivered using a local offline algorithm, which takes into account all currently available information. Such an approach can be viewed as aiming to maximize the benefit to be accrued from the present packets, assuming no future arrivals. Such an approach lays at the core of the solutions proposed by~\cite{kesselman04buffer,jez12online,englert12considering} which each used an algorithm for computing an optimal offline local solution.
In our case, as shown in \autoref{cor:offline:hardness}, computing such an optimal provisional schedule is hard, but, as shown in \autoref{cor:offline:alg}, there exists a $(k+1)$-approximation algorithm for the problem.

We adapt this algorithm into a procedure for computing a provisional schedule, which would allow a smaller $I$-indexed frame to have one of its packets scheduled, only if non of the frames with a higher $I$-index would become infeasible in the following time slot.
Our proposed heuristic, \oplong, is described in \autoref{alg:opportunistic}. \oplong\ builds a provisional schedule $F_t$ as follows:
\begin{enumerate}
\item Sort pending frames\footnote{A frame is pending if it has pending packets.} in decreasing lexicographical order of $(I_t(f),d-r_t(f))$. I.e., preference is given to frames with higher $I$-index values. In case of ties,
preference is given to frames for which their smallest-index packet has the minimal residual slack.
\item Initialize the provisional schedule $F_t=\emptyset$.
\item For each frame $f$ in this order, test whether for all $s=0,\ldots,d$, the pending packets of $f$ can be added to $F_t$ such that the overall number of packets in the provisional schedule with remaining slack at most $s$, does not exceed $s$.
    If $f$ can be added, update $F_t=F_t \cup \set{f}$.
\end{enumerate}

\begin{figure}[htb]
\centering
\includegraphics[width=9cm]{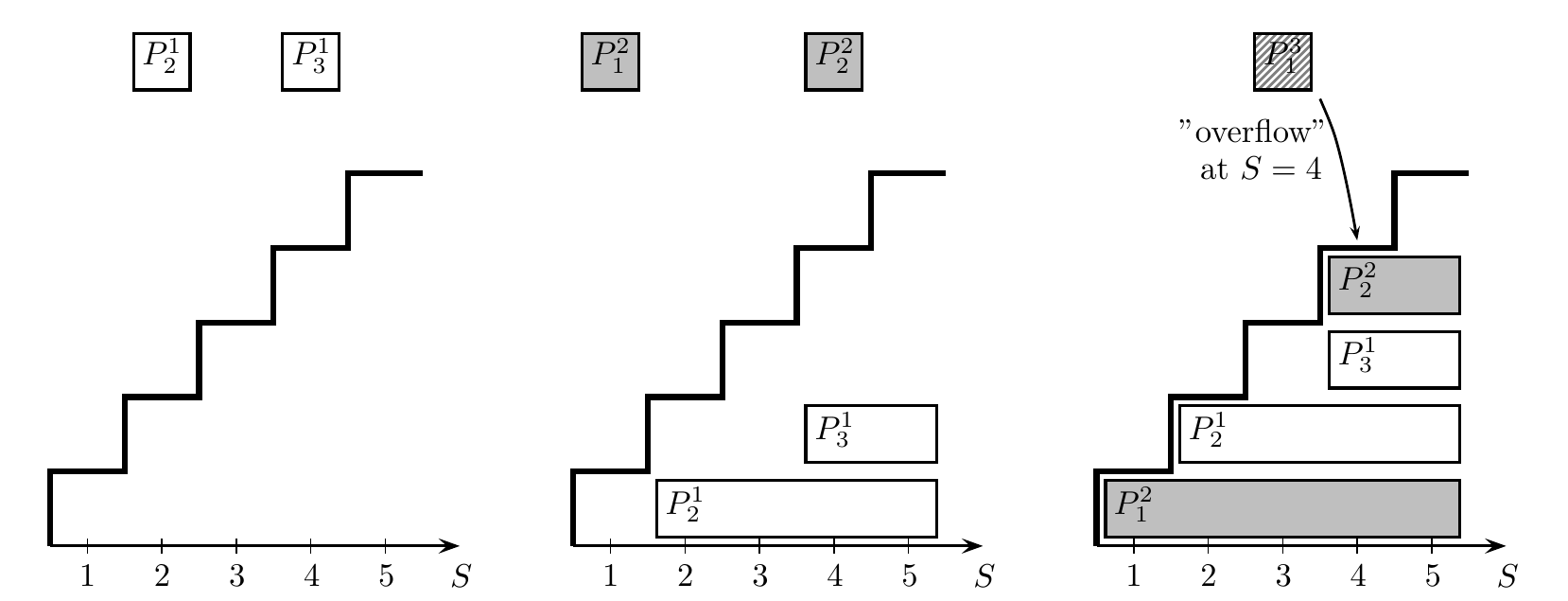}
\caption{Schematics of building a provisional schedule.}
\label{fig:prov-sched}
\end{figure}

\autoref{fig:prov-sched} gives an illustration of construction of a provisional schedule at some time $t$.
The first two frames have room for all their packets in the provisional schedule, and all of their packets can be accommodated for delivery by their deadlines (note that the first frame tested will always be a part of the provisional schedule, since otherwise it would not be alive).
The packet of the third frame causes an ``overflow'' for $s=5$, and therefore its frame cannot be accommodated in the provisional schedule.
We note that in this example, the packet picked for delivery would be $P_1^2 \in f_2$, which might not correspond to the highest $I$-index pending frame (e.g., if $I_t(f_2)<I_t(f_1)$).
%

\begin{algorithm}{}
\caption{\oplong: at the scheduling substep of time $t$}
\label{alg:opportunistic}
\begin{algorithmic}[1]
\State Build the provisional schedule $F_t$
\State transmit the packet with minimum residual slack in $F_t$
\end{algorithmic}
\end{algorithm}

\section{Simulations}
\label{sec:simulations}

In the previous sections we provided an analysis of the greedy online algorithm, and of discussed additional guidelines for effective algorithm design. We also presented algorithm \oplong.
In this section we provide a simulation study in which we test the performance of the algorithms, the effectiveness the guidelines, and the impact the parameters $k$ and $d$ on the performance.

\subsection{Traffic Generation and Setup}

We recall that the problem of managing traffic with packet dependencies captured by our model is most prevalent in real-time video streams.
We therefore perform a simulation study that aims to capture many of the characteristics of such streams.

We will generate traffic which will be an interleaving of {\em streams}, where each stream is targeted at a different receiver, and all streams require service from a single queue at the tail of a link.

%
In our simulation study, we focus on traffic with the following characteristics.
\begin{itemize}
\item We assume each stream has a random start time where packets are generated. This corresponds to scenarios such as VOD streams, where each receiver may choose which video to view and when to view it, and these choices are independent.
\item We assume the average bandwidth demand of all streams is identical, which represents streams with comparable video quality.
\item Frames of a single stream are non-overlapping, and are produced by the source in evenly spaced intervals. E.g., if we consider video streams consisting of 30FPS, each interval is $33ms$.
\item The source transmits the packets of each frame in a burst, and we assume each frame consists of the same number of packets. Such a scenario occurs, e.g., in MPEG encoding making use of I-frames alone. We assume all packet have the same size, namely, the network's MTU.
\item We assume a random delay variation between the arrival of consecutive packets corresponding to the same stream. Such delay variation is produced, e.g., due to queueing delay in previous nodes along the streams path. Specifically, we assume a uniform delay variation of up to 5 time slots.
\item We assume each packet contains the frame number, and the index number of the packet within the frame. Such information can be encoded, e.g., in the RTP header.
\end{itemize}

For the setup we chose to simulate, the packet sizes are set such that every time slot one packet can be scheduled, and the aggregate bandwidth of all the streams is equal to the service bandwidth. We note that even in such cases, where traffic arrival rate does not exceed the link capacity, no online algorithm can obtain the optimal goodput.

We simulate 2 minutes worth of traffic for 50 streams, where for all the streams the frame rate is 30FPS (for a total of 3600 frames per stream). Since we fix the service rate as $1$, in the simulation the interval between consecutive frames arrival in a stream is $\Delta F=k\cdot 50$ time slots, where $k$ is the number of packets per frame. As $k$ grows the ``real'' duration of a single time slot decreases, as the service rate effectively increases.

\subsection{Simulated Algorithms}

We performed the simulation study for four scheduling algorithms, where in all algorithms in case of ties in the priorities, these are broken according to the a random (but fixed) priority on the streams:
\begin{itemize}
\item The offline $O(k+1)$-approximation algorithm of~\cite{kesselman13competitive}. By \autoref{cor:offline:alg} this algorithm has the same performance guarantee in our model as well. This algorithm serves as a benchmark for evaluating the performance of the online algorithms.
\item Algorithm \gdlong, described and analysed in \autoref{sec:greedy}. This algorithm represents our baseline for studying the the performance of online algorithms for the problem.
\item Algorithm \gdslong, which implements \gdlong\ with ties broken according to minimum residual slack.
\item Algorithm \oplong, presented in \autoref{sec:algorithm-design:scheduling}.
    This is the most complex algorithm we evaluate, as in addition to the enhanced tie-breaking, it also attempts to exploit opportunities to schedule lower ranked packets according to the provisional schedule.
\end{itemize}

\subsection{Results}

The simulation results confirm our hypothesis that implementation of the proposed algorithm design guideline does indeed impact the performance of online algorithms.
We depict the performance of each online algorithm by its {\em goodput ratio}, measured by the ratio between the goodput of the online algorithm and that of the offline algorithm.

\begin{figure}[htp]
\centering
\includegraphics[width=9cm]{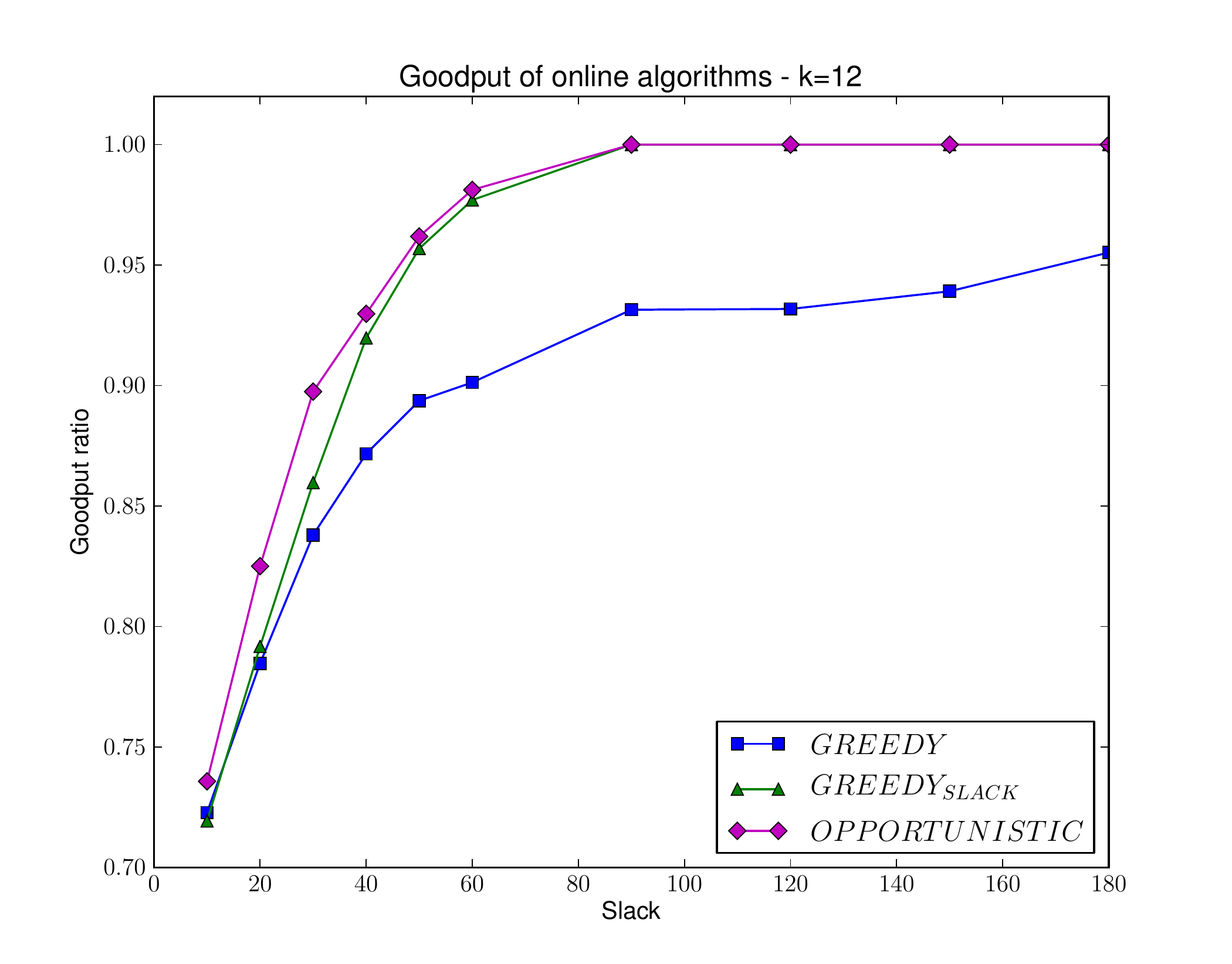}
\caption{Comparison of the goodput of online algorithms}
\label{fig:comp-on-alg}
\end{figure}

\autoref{fig:comp-on-alg} presents the performance of the online algorithms as a function of the slack each packet has, for $k=6$. It can be seen that as the slack increases the tie-breaking rule in \gdslong\ shows significant improved performance in comparison with the vanilla greedy algorithm.
The figure also shows that the \oplong\ exhibits a significantly better performance than \gdslong\ (although this improvement is paid for by significant additional complexity).
We note that results for greater values of $12$ exhibit the same trends.
Also of note is that the \gdslong\ and \oplong\ manage to trace the performance of the offline algorithm (and actually complete all the frames of all the streams) for traffic with sufficiently large slack.

\begin{figure*}[htp]
\centering
\subfloat[\oplong]{\includegraphics[width=6cm]{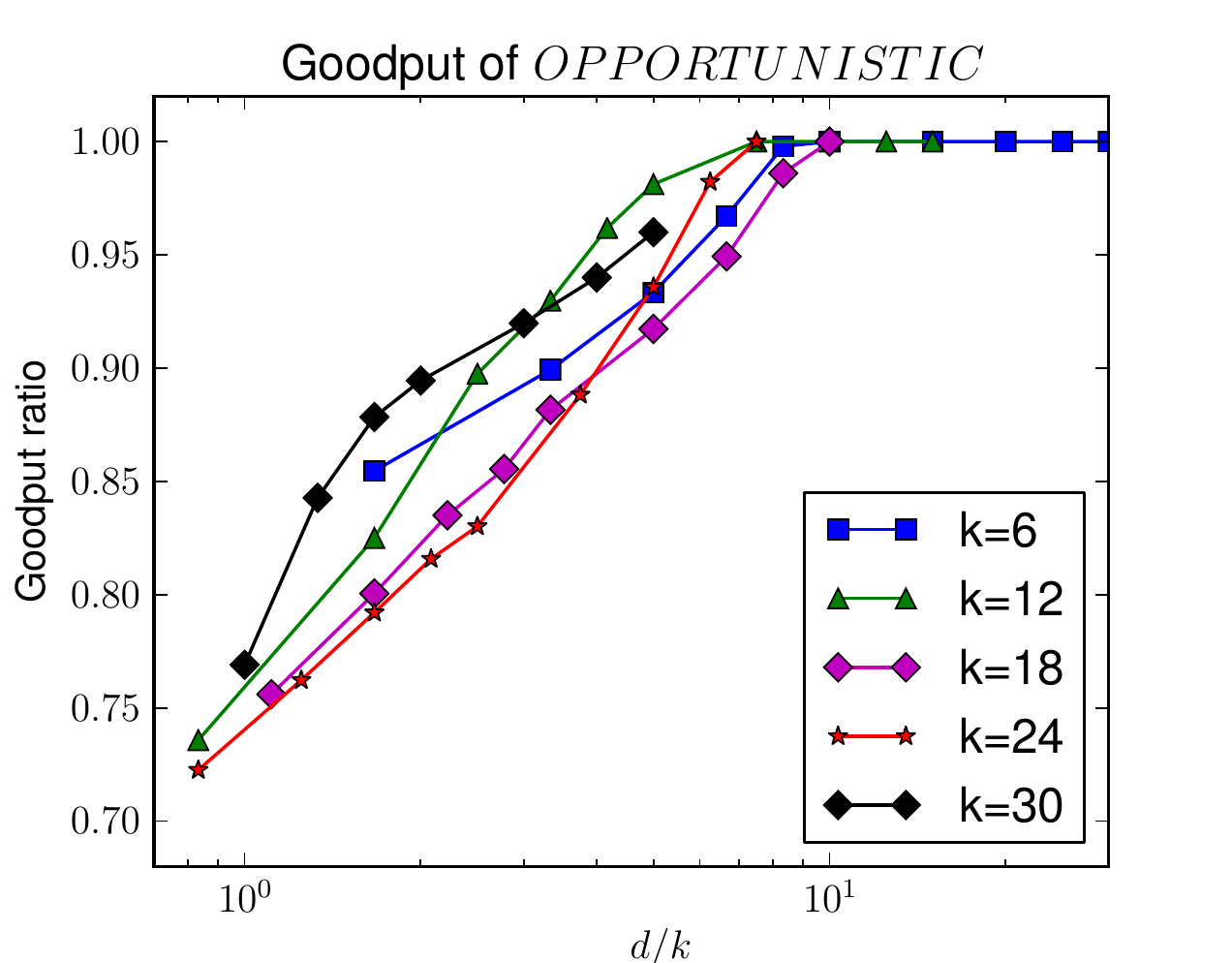}}
\subfloat[\gdslong]{\includegraphics[width=6cm]{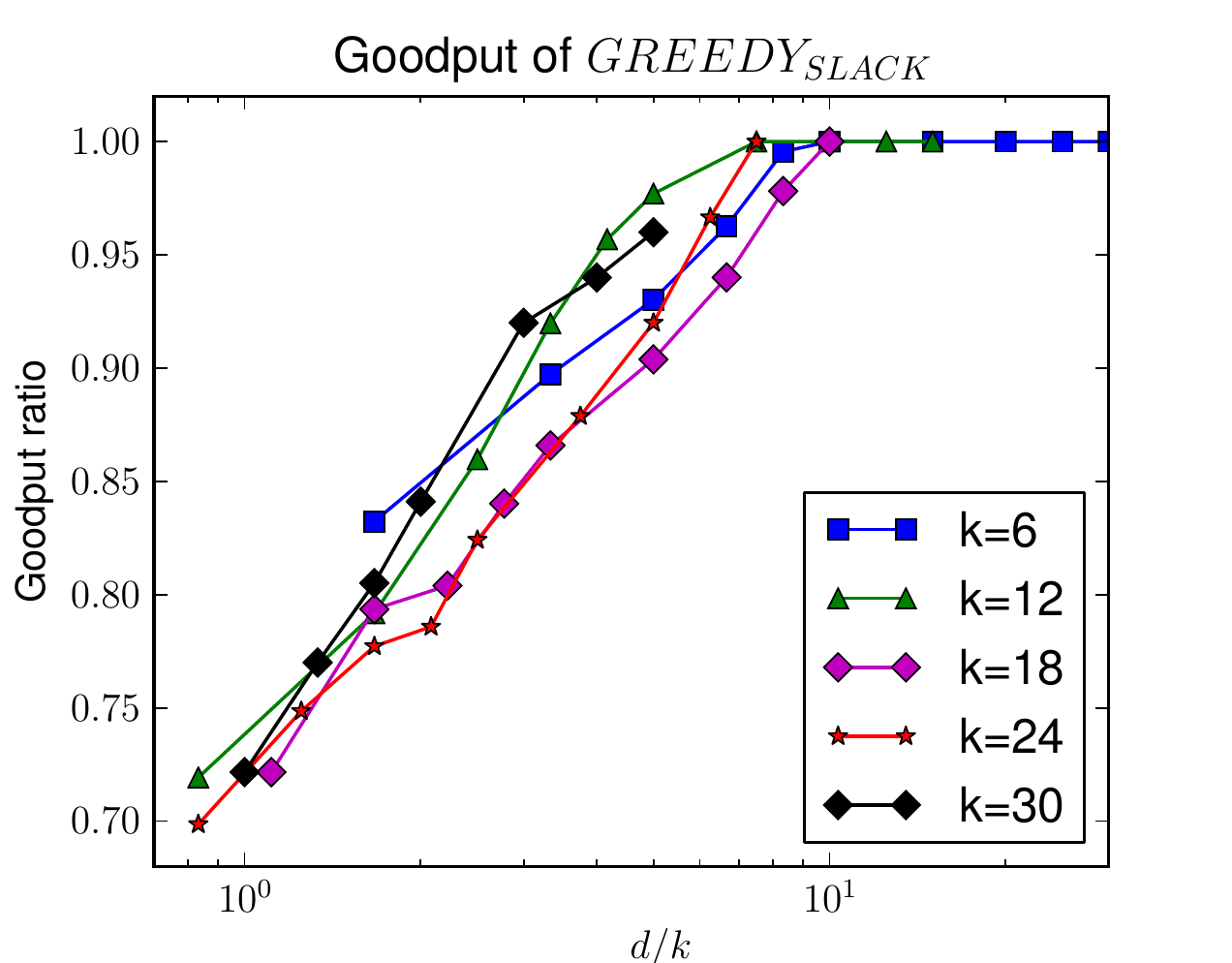}}
\subfloat[\gdlong]{\includegraphics[width=6cm]{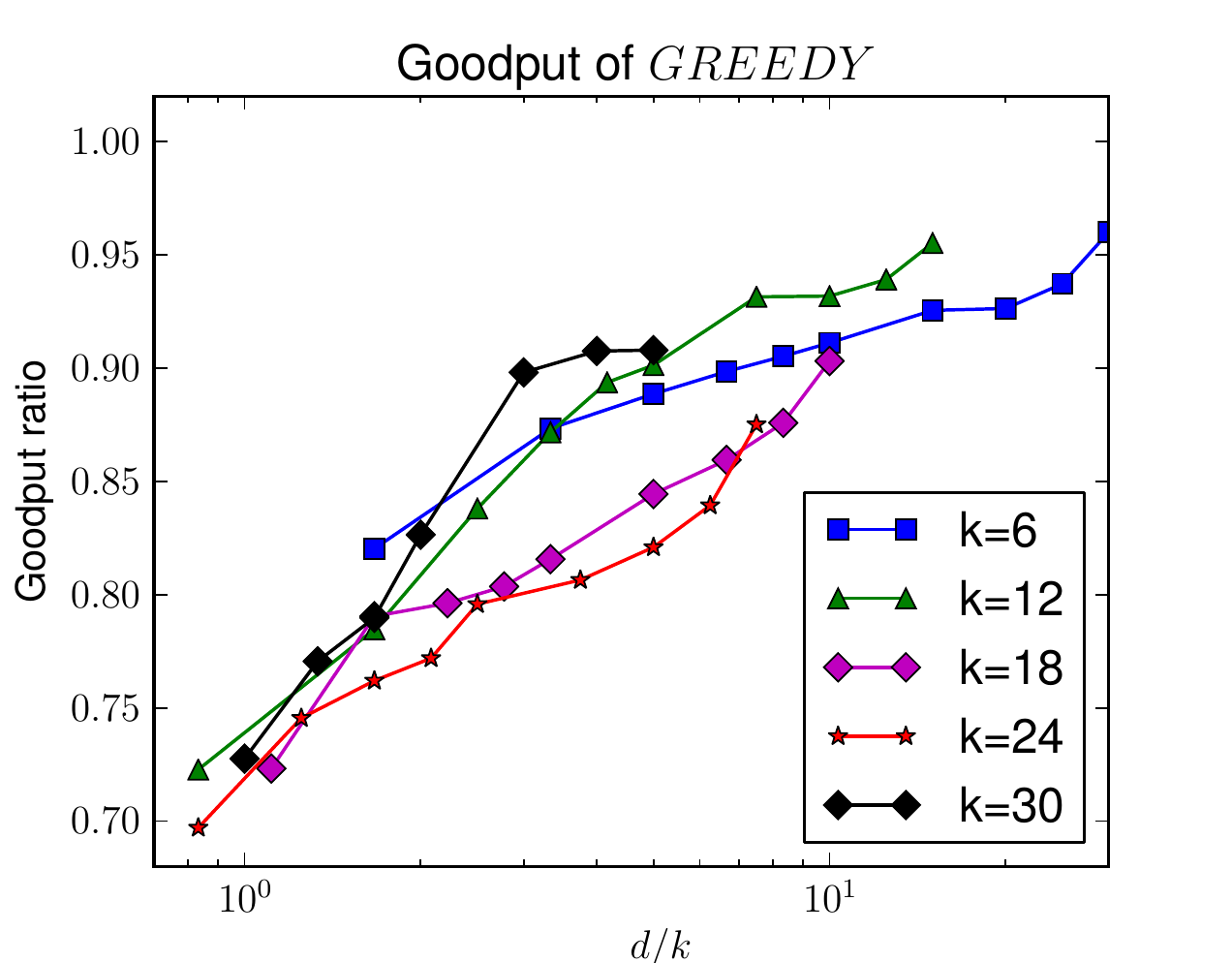}}
\caption{Goodput of the online algorithms as a function of $d/k$ on a logarithmic scale}
\label{fig:algorithms-dk}
\end{figure*}


In \autoref{fig:algorithms-dk} we presents the goodput ratio of \oplong\ and \gdslong\ as a function of $d/k$.
The first lesson learnt from this data is that the performance of the opportunistic algorithm is superior to that of the enhanced greedy algorithm, in particular for small $d/k$ values where the difference becomes more pronounced (these results are also hinted by \autoref{fig:comp-on-alg}, but are not as pronounced).
Furthermore, the performance of both algorithms depends exponentially on the ratio between $d$ and $k$ (shown by the log scale), as even though both graphs present results of many simulations with different inputs having different parameters, the plots show a linear trend up to the point where they match the goodput of offline algorithm.
This exponential dependency can be viewed as comparable to that of the analytic lower bound presented in \autoref{sec:online:lower-bounds}, where $d/k$ takes the role of $1/k$.

\begin{figure}[htp]
\centering
\includegraphics[width=9cm]{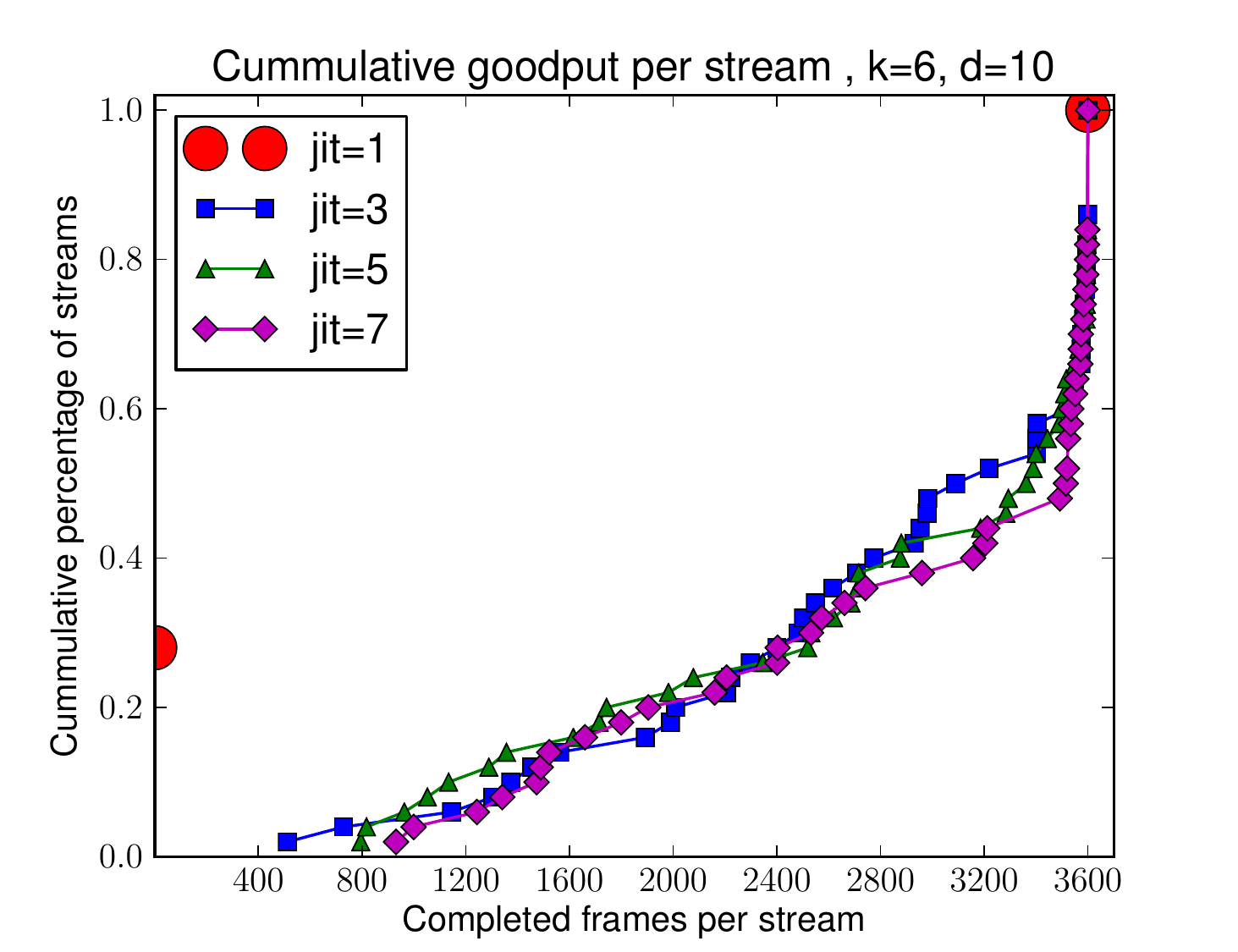}
\caption{Cumulative completed frames per stream for \gdslong\ as a function of maximal delay variation (jitter) between packets of a stream.}
\label{fig:cdf-jit}
\end{figure}

Another important result is that the streams are not treated fairly.
When there is no random delay variation for the input sequence, the algorithms synchronize with the input so either all frames of a stream are completed, or none of them are completed. Raising the limits of the random delay variation improves the fairness, although some degree of synchronization remains. When there exists a random delay variation higher values of $k$ result in improved fairness. The impact of the delay variation can be seen in \autoref{fig:cdf-jit}, which shows that for \gdslong\ maximal delay variation of 1 time slot (for an average of 0.5 time slots) results in complete synchronization - a stream either has all of it's frames completed or it has no frame completed. Raising the maximal delay variation even by a small amount reduces the synchronization significantly. 

\section{Conclusions and Future Work}
\label{sec:conclusions-future-work}

In this paper we address the problem of maximizing the goodput of delay sensitive traffic
with inter-packet dependencies.
We provide lower bounds on the competitiveness of online algorithms for the general case that the traffic is burst bounded, and present competitive scheduling algorithms for the problem.
%
Through the analysis we show that there exists an algorithmic guideline that ensures competitiveness - preference for frames that are closer to completion.
Our proposed solutions ensure the optimal performance possible, up to a small constant factor.

Our analysis further provides insights into improving the performance of online algorithms for the problem. These insights are further verified by a simulation study which shows that our improved algorithms which are inspired by our analytic results, are very close to the performance of the currently best known offline algorithm for the problem. More specifically, the performance of our algorithms approach the performance of our benchmark algorithm with an exponential correlation to the increase in delay-slack.

Our work serves as an initial study of scheduling delay-bounded traffic with inter-packet dependencies.
Our work raises new questions about the performance of algorithms for this problem:
\begin{enumerate}
\item Our simulation results indicate that the ratio $d/k$ bears some influence on the algorithm performance. Shedding light on this effect is an interesting open question.
\item Are there other algorithmic guidelines which can further improve the performance of online algorithms, and specifically how well can randomized algorithms perform?
\end{enumerate}

\bibliographystyle{IEEEtran}
\bibliography{MS}

\end{document}